%% file: Document/arxiv_main.tex
\documentclass[letterpaper]{article}

\usepackage{fancyhdr}

\usepackage{xcolor}

\usepackage{graphicx}

\usepackage{gensymb}
\usepackage{amssymb}
\usepackage{amsmath}

\makeatletter 
\newif\if@anonymous 
\@anonymousfalse 

\usepackage[colorlinks=true, allcolors=blue]{hyperref}

\usepackage[left=2.75cm,right=2.75cm,top=2.75cm,bottom=2.75cm]{geometry}
\newcommand{\articletype}[1]{}

\renewcommand{\title}[1]{{\exhyphenpenalty=10000\hyphenpenalty=10000 
 \fontsize{18}{21}\selectfont\noindent\raggedright
        \textsf{#1}\par}\suppressfloats[t]}

\renewcommand{\author}[1]{{\vspace{5mm}%
   \fontsize{10}{12}
      \raggedright \if@anonymous Author list removed for anonymity \else #1 \fi
	  \vspace{3mm}}}

\newcommand{\affil}[1]{{\fontsize{8}{10}\selectfont
       \raggedright \if@anonymous \phantom{#1} \else #1 \fi}
	   }

\newcommand{\email}[1]{\vspace*{12pt}{\fontsize{8}{10}\selectfont
       \raggedright {\bfseries E-mail:} \if@anonymous \phantom{#1} \else #1 \fi}
	  \vspace{3mm} }
	   
\newcommand{\keywords}[1]{{\fontsize{8}{10}\selectfont
       \raggedright {\bfseries Keywords:} #1}
	  }

\renewcommand\section{\@startsection {section}{1}{\z@}%
                   {-3.25ex\@plus -1ex \@minus -.2ex}%
                   {1sp}%
                   {\reset@font\normalsize\bfseries\raggedright}}
\renewcommand\subsection{\@startsection{subsection}{2}{\z@}%
                   {-3.25ex\@plus -1ex \@minus -.2ex}%
                   {1sp}%
                   {\reset@font\normalsize\itshape\raggedright}}
\renewcommand\subsubsection{\@startsection{subsubsection}{3}{\z@}%
                                     {-3.25ex\@plus -1ex \@minus -.2ex}%
                                     {-1em \@plus .2em}%
                                     {\reset@font\normalsize\itshape}}

\newcommand{\ack}[1]{
\section*{Acknowledgments}
\if@anonymous Removed for anonymity \else #1 \fi}

\newcommand{\funding}[1]{
\section*{Funding}
\if@anonymous Removed for anonymity \else #1 \fi}

\newcommand{\coi}[1]{
\section*{Conflict of interest}
\if@anonymous Removed for anonymity \else #1 \fi}

\newcommand{\data}[1]{
\section*{Data availability}
\if@anonymous Removed for anonymity \else #1 \fi}

\newcommand{\roles}[1]{
\section*{Author contributions}
\if@anonymous Removed for anonymity \else #1 \fi}

\newcommand{\suppdata}[1]{
\section*{Appendix}
#1}

\renewcommand{\@makecaption}[2]{\vskip\abovecaptionskip
\sbox\@tempboxa{\fontsize{8}{10}\selectfont {\bfseries #1.} #2}%
\ifdim \wd\@tempboxa >\hsize
\raggedright \fontsize{8}{10}\selectfont {\bfseries #1.} #2\par
\else
\global \@minipagefalse
\hb@xt@\hsize{\hfil\box\@tempboxa\hfil}%
\fi
\vskip\belowcaptionskip}

\let\oldtabular\tabular
\renewcommand{\tabular}{\fontsize{8}{10}\selectfont \oldtabular}

\usepackage{orcidlink} 

\usepackage{harvard} 
\citationmode{abbr} 

\makeatother

\begin{document}

\articletype{Paper} 

\title{Non-circular scan trajectories for reducing cone-beam artifacts in Gamma Knife CBCT images: a simulation study}

\author{Alexandra Alain-Beaudoin$^{1,2}$\orcidlink{0009-0001-7599-8448}, Håkan Nordström$^3$\orcidlink{0000-0003-3751-7700}, Luc Beaulieu$^{1,2}$\orcidlink{0000-0003-0429-6366} and Joakim da Silva$^{3,*}$\orcidlink{0000-0002-7771-842X}}

\affil{$^1$Département de physique, de génie physique et d'optique, et Centre de recherche sur le cancer, Université Laval, Québec, Canada}

\affil{$^2$Service de physique médicale et radio-protection, et Axe Oncologie du CRCHU de Québec, CHU de Québec Université Laval, Québec, Canada}

\affil{$^3$Physics and Advanced Applications, Elekta Instrument AB, Stockholm, Sweden}

\affil{$^*$Author to whom any correspondence should be addressed.}

\email{joakim.dasilva@elekta.com}

\keywords{Gamma Knife, cone-beam computed tomography (CBCT), cone-beam artifacts, sampling incompleteness, scan trajectory, CBCT simulations}

\input{Document/Abstract}

\input{Document/Intro}

\input{Document/Method}

\input{Document/Resultats}

\input{Document/Discussion}

\input{Document/Conclu}

%
%


\funding{This study was supported by Mitacs through the Mitacs Accelerate International program [reference number IT42919]. In this context, Elekta Instrument AB (Stockholm, Sweden) partly funded the study. 
This work was also supported by the Natural Sciences and Engineering Research Council of Canada [funding reference number 596759-2024]; and the Fonds de recherche du Québec [https://doi.org/10.69777/351806].}

\coi{Håkan Nordström and Joakim da Silva are employees of Elekta Instrument AB. Alexandra Alain-Beaudoin receives a stipend partly funded by Elekta Instrument AB. Luc Beaulieu has no conflict of interest to disclose.}


\roles{
Alexandra Alain-Beaudoin: Data curation, formal analysis, funding acquisition, investigation, methodology, software, validation, visualization, writing - original draft, writing - review \& editing
\newline
Håkan Nordström: Conceptualization, resources, writing - review \& editing
\newline
Luc Beaulieu: Funding acquisition, project administration, supervision, writing - review \& editing
\newline
Joakim da Silva: Conceptualization, methodology, project administration, resources, supervision, validation, writing - review \& editing}



\suppdata{
\input{Document/Annexe}
}

\clearpage


\input{Document/bbl.tex}
\end{document}

%% file: Document/Abstract.tex
\section*{Abstract}

\textit{Objective.} Gamma Knife cone-beam computed tomography (CBCT) images are deteriorated by cone-beam artifacts whose magnitude increases along the superior direction. In this study, a novel scan trajectory compatible with the Gamma Knife CBCT system is optimized to reduce cone-beam artifacts, with the aim to replace the current 200-degree single-arc scan. 
\textit{Approach.} Data sampling analysis with tomographic incompleteness maps indicates the level of undersampling across the field of view based on the geometry of the system and of a scan trajectory. Moreover, CBCT simulations are performed from a virtual phantom with disks aligned along the axial direction and from a CT reconstruction of a stereotactic end-to-end validation (STEEV) phantom. CBCT projections are simulated for a given scan trajectory through a polychromatic forward projection model with added noise and scatter, then the CBCT image is reconstructed using an iterative algorithm which minimizes weighted least squares.
\textit{Main results.}
Both the incompleteness analysis and the CBCT simulations indicate adding lines to the current single-arc trajectory is more efficient to reduce cone-beam artifacts than adding more arcs, both in terms of number of additional projections and new artifacts. A line-arc-line trajectory with source axial steps of 3.5 mm removes virtually all cone-beam artifacts. The widths of the cone-beam artifacts created by the disks show a positive correlation between the artifact magnitude and the incompleteness value.
\textit{Significance.}
A line-arc-line scan trajectory is promising to reduce cone-beam artifacts of the Gamma Knife CBCT images while being a compatible and reasonable trajectory for the current system design.

%% file: Document/Intro.tex
\section{Introduction}

Radiosurgery requires high accuracy and precision in treatment delivery. In Gamma Knife treatments, patient positioning and monitoring is achieved partly through the integrated cone-beam computed tomography (CBCT) system \cite{Zeverino_Icon_CBCT}. However, its images often have noticeable cone-beam artifacts towards the top of the head, degrading the image quality. These artifacts are caused by the circular arc scan trajectory and the geometric limitations of the CBCT system imposed by the Gamma Knife design.

The CBCT system has the piercing point located at the inferior edge of the detector, creating large cone-beam angles towards the superior side of the detector, as shown in Figure \ref{fig: schema cbct + phantom}. When using a single-arc scan trajectory, the CBCT images are degraded by visible cone-beam artifacts. While the presence of cone-beam artifacts is object-dependent, their magnitude increases with the cone angle. In practice, in a head reconstruction, the superior side of the skull sees the largest artifact \cite{Zeverino_Icon_CBCT}.

To accurately reconstruct a point in the field of view (FOV), Tuy’s local sufficiency condition for untruncated projections states that every plane containing the point must intersect the source trajectory \cite{tuy_inversion_1983}. In the idealized context of a single-circle scan with continuous source trajectory, Tuy’s condition implies that only the trajectory plane is fully sampled, and constituently allows artifact-free reconstruction. However, Tuy’s condition is not adapted to the discretized source positions of a realistic scan and to the finite size of the detector which may lead to truncated projections. Moreover, Tuy’s condition only establishes if a point in the FOV is fully sampled, it does not quantify the level of sampling completeness, or incompleteness for incompletely sampled points. While quantifying completeness can be complex and unreliable without any prior information about the object and system configurations, an incompleteness metric can be established based on Finch’s statement that, if a point has an intersecting plane which crosses no X-ray source location, stable reconstruction at that point is impossible \cite{finch_cone_1985}. Therefore, a metric for sampling incompleteness can be defined based on the distance to the closest source location for a given plane. 

\citeasnoun{clackdoyle_quantification_2019} have proposed an incompleteness metric that indicates the severity of artifacts which may appear in the reconstructed volume at a point because of missing information in a given plane direction, regardless of the reconstruction algorithm used. \citeasnoun{laurendeau_three-dimensional_2023} generalized this concept to reach an overall incompleteness metric that considers every plane intersecting a given point to output a unique incompleteness value. This incompleteness metric can be employed to evaluate and compare different scan trajectories for a CBCT system. 

Many publications, reviewed by \citeasnoun{hatamikia_source-detector_2022}, propose more complete scan trajectories in terms of sampling compared to the circular scan trajectory. In recent years, for single source scans, attention has been centered around saddle trajectories \cite{wei_reduction_2024,cancelliere_butterfly_2023,hosoo_image_2023,jones_cone-beam_2024} as well as ellipse-line-ellipse or line-ellipse-line trajectories \cite{yu_extended_2016,guo_c-arm_2020,yu_line_2011}. Saddle trajectories, also called butterfly or sine-on-sphere scans, are circular scans with a simultaneous sinusoidal motion of the gantry tilt angle. Their sampling can fulfill Tuy’s condition, allowing exact reconstruction. However, adding a gantry tilt is not possible for the Gamma Knife CBCT system due to mechanical constraints. Next, ellipse trajectories are arcs with a simultaneous short translation of the source, creating an ellipse in the tilted plane of the trajectory. Compared to the single-arc scan, using a combination of lines and ellipses as trajectory increases the axial coverage while improving the sampling completeness. This type of trajectory is advantageous for standard CBCT designs where the piercing point is centered on the detector, but not for the Gamma Knife CBCT. As the Gamma Knife's piercing point is on the inferior edge of the detector, it is expected that a circular arc must remain on the inferior edge of the object. Arcs with simultaneous source translations are referred to as helices in the rest of this work.

The goal of this work is to find a new scan trajectory compatible with the Gamma Knife mechanical design, which improves sampling and reduces cone-beam artifacts, while keeping in mind the practical feasibility of the trajectory. To this end, tomographic incompleteness maps are computed for various scan trajectories, allowing theoretical analysis of the undersampled regions of the FOV which may lead to cone-beam artifacts. Using a virtual phantom with disks aligned along the axial axis and a CT reconstruction of a head phantom, simulations of the CBCT system are realized to assess the new potential scan trajectories. The results from the theoretical analysis and the simulations are compared to reach a conclusion about the best scan trajectory for the Gamma Knife CBCT system.

%% file: Document/Method.tex
\section{Methods}

\subsection{Scan Geometry and Trajectories}

\begin{figure}[htb!]
    \centering
    \includegraphics[width=1\linewidth]{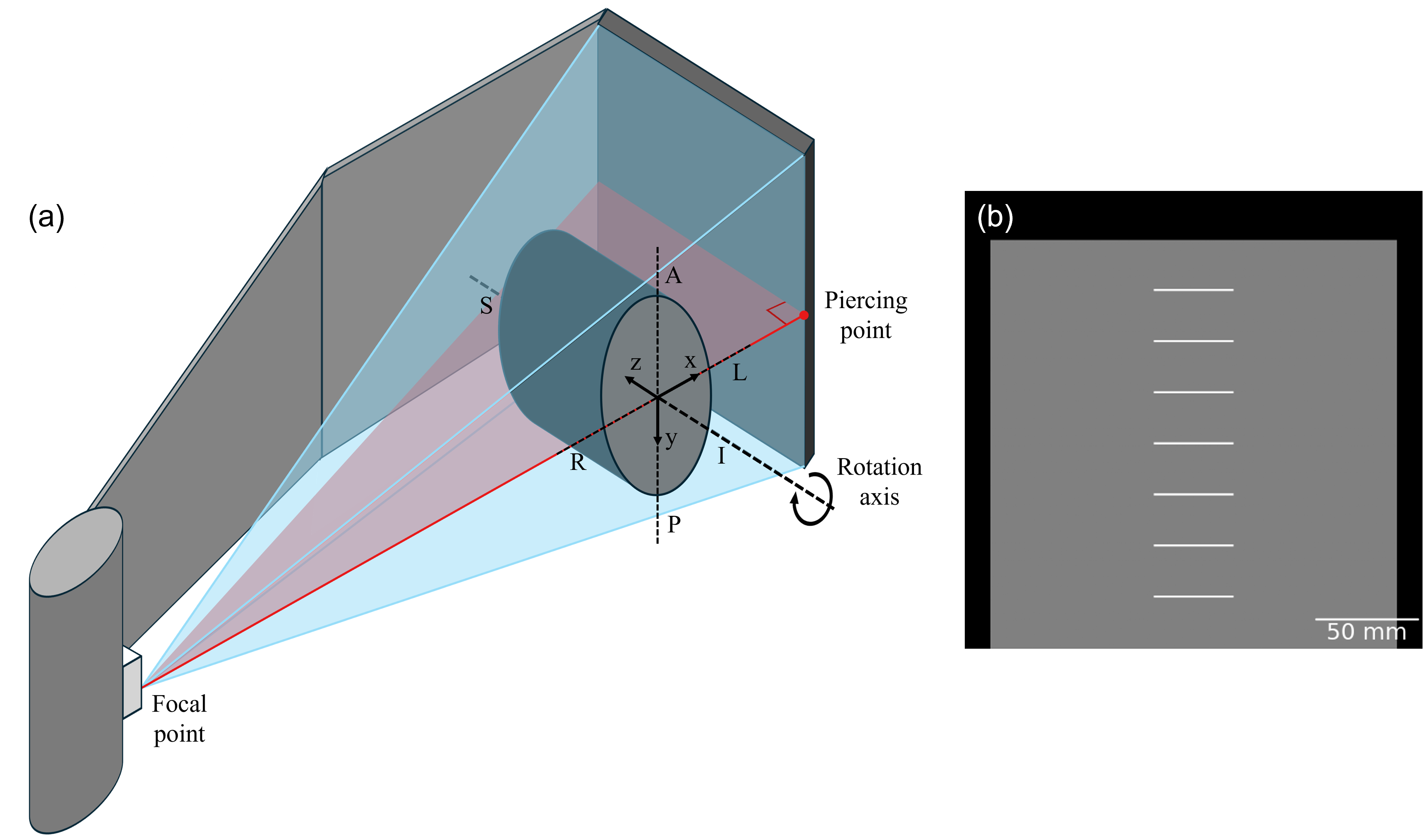}
    \caption{Gamma Knife CBCT geometry in (a), with the imaging coordinate system and the anatomical directions, followed by the central coronal slice of the virtual phantom with disks aligned along the axial direction in (b).}
    \label{fig: schema cbct + phantom}
\end{figure}

The geometry of the Gamma Knife CBCT is depicted in Figure \ref{fig: schema cbct + phantom}(a). The imaging coordinate system is fixed in space and the gantry rotates around the z-axis during the arc scan of 200$\degree$. At the inferior (I) edge of the scanned object, the X-ray beam is perpendicular to the detector plane. Towards the superior (S) direction, the beam becomes increasingly inclined compared to the detector, leading to increasingly larger cone-beam artifacts in the reconstruction.

The Gamma Knife CBCT design restricts the possible degrees of freedom for creating a new scan trajectory to moving the patient table in three dimensions and rotating the source around the z-axis. The scan trajectories possible are limited to combinations of arcs at different z-positions of the table, lines by moving the table along the z-axis for the available gantry angles, and helices by simultaneously moving the table and rotating the source. For practical reasons, the configurations investigated for the lines are limited to the first and last gantry angles of the 200$\degree$ arc. Lines up to 175 mm in range are investigated, based on the typical size of a scanned object and on an assessment of the Gamma Knife mechanical limitations. The axial step of the source is varied between 7 mm and 1.75 mm, leading to line scans made of 25 to 100 projections, while arcs and helices are made of 334 projections, like the current single-arc trajectory. The scan trajectories investigated are simulated using the exact dimensions of the Gamma Knife system.

\subsection{Tomographic Incompleteness}

To quantify the amount of information missing for exact reconstruction, \citeasnoun{clackdoyle_quantification_2019} proposed a local directional incompleteness metric $I(\textbf{p}, \textbf{n}) \in \mathbb{R}^+$ at a point of interest $\textbf{p} \in \mathbb{R}^3$ for a direction $\textbf{n} \in S^2$ on a unit sphere. Defining $\psi_i$ as the angle between the line defined by \textbf{p} and a source location $\textbf{s}_i \in \{\textbf{s}_1,\textbf{s}_2,…,\textbf{s}_N \} \in \mathbb{R}^{3 \times N}$, and the plane of normal direction $\textbf{n}$ passing through $\textbf{p}$, the local directional tomographic incompleteness is expressed as 
\begin{equation}
    I(\textbf{p}, \textbf{n}) = \min_{i \in \{1,2,...,N\}} {\tan \psi_i}
\end{equation}
The directional tomographic incompleteness at \textbf{p} and \textbf{n} represents the minimum tangent of the angles  $\psi_i$ defined by all source positions, where $\psi$ is the angle of the closest source position. If $I(\textbf{p}, \textbf{n}) = 0$, the plane intersects with a source position. As $I(\textbf{p}, \textbf{n})$ increases, so does the data insufficiency, and the farther the case is from satisfying Tuy's condition.
As proposed by \citeasnoun{laurendeau_three-dimensional_2023}, the incompleteness $I_{\infty}$ at each point $\textbf{p}$ of the scanned volume can be defined as the maximum incompleteness value for every normal direction \textbf{n}.
\begin{equation}
    I_{\infty}(\textbf{p}) = \max_{\textbf{n} \in S^2} { I(\textbf{p}, \textbf{n})}
\end{equation}
\citeasnoun{sun_2017} and \citeasnoun{wei_reduction_2024} have reported that no artifacts are visible if $I_{\infty} < 0.02$, which means data sampling is sufficient for reconstruction without cone-beam artifacts.

Moreover, a beam intersecting a point \textbf{p} must reach the detector to provide new data for the reconstruction. Therefore, to achieve a more realistic assessment of the sampling incompleteness, only the source positions which project \textbf{p} onto the detector are considered. This realistic incompleteness definition, which is used from this point onward, leads to a $I_{\infty}$ value that is expected to be the same as, or higher than, that for an infinite detector size. 

The tomographic incompleteness $I_{\infty}$ is used to assess the potential of different scan trajectories to produce artifact-free reconstructions through improved data sampling.
$I_{\infty}$ is computed across the scanned volume region of the Gamma Knife CBCT; more specifically at every 5-mm axial step, for 37 points distributed across the axial plane of the FOV with a 112-mm radius.
For the normal directions \textbf{n}, 200,001 points are approximately uniformly sampled on a hemisphere using a Fibonacci lattice \cite{fibonacci_2010}. 
Assuming equidistance between points on the Fibonacci lattice, an upper bound for the estimated incompleteness values is estimated using half the cone angle between the points. 
The incompleteness maps of each scan trajectory investigated are evaluated based on the overall distribution of the estimated $I_{\infty}$ in the FOV, and with metrics such as the median and the 90th percentile of the distributions.

\subsection{CBCT Simulations}

\subsubsection{Phantoms}

A virtual phantom, shown in Figure \ref{fig: schema cbct + phantom}(b) is created to characterize the magnitude of cone-beam artifacts across the volume. This cylindrical phantom, made of soft tissue-equivalent material, contains 1 mm-thick disks of 40-mm diameter spaced 25 mm apart, arranged along the axial direction and centered on the z-axis. The disks are made of bone-equivalent material, creating a high-density variation. 
To evaluate the artifacts outside the z-axis center, more virtual phantoms are created with the disks positioned at $\pm60$ mm along each axis in the axial plane.
The effect of the geometry of the disks on the magnitude of the cone-beam artifacts is studied with the addition of other virtual phantoms with disk thicknesses of 2 mm and with diameters of 20 and 30 mm.

A CT reconstruction of a Stereotactic End-to-End Verification (STEEV) head phantom (Computerized Imaging Reference Systems, Inc., Norfolk, United States), shown in Figure~\ref{fig: steev best vs arc wls vs ct}(a,d), is also used for simulations of a typical object scanned on a Gamma Knife. The volume is segmented into three materials: air, soft tissue and bone. Densities are assigned based on a piece-wise linear function of the Hounsfield values of the CT volume.

\subsubsection{Software}

CBCT images for different scan trajectories are generated through a simulator separated into two parts: a CBCT projection simulator and a reconstruction software.

The projection simulator is a research software developed at Elekta, which produces a CBCT projection stack based on an object volume input, a system geometry, beam and detector models, and a scan trajectory. The simulator is based on a polychromatic forward projection model, which computes an energy integral using the spectrum and the object’s attenuation, the latter based on the accumulated density of all materials in the volume. Quantum and electronic noise are also added during the simulation. The spectrum and beam profile are obtained via a Monte Carlo-generated phase space of the Gamma Knife X-ray beam. Apart from the primary signal, scatter is also simulated using GPU Monte Carlo simulation for Imaging (GPUMCI)~\cite{gpumci_2017}, which requires the beam’s phase space as well as attenuation and interaction tables for each material (see TG-268 checklist at Table \ref{tab: tg268} of the Appendix). 

Once a projection stack is acquired for a specific scan trajectory, the reconstruction software is used to obtain a CBCT volume. 
The reconstruction is made with an iterative algorithm from Elekta based on minimizing weighted least squares (WLS) \cite{book_wls} using the Fast Gradient Method (FGM)~\cite{FGM}, with pseudo-Huber Total Variation (TV) regularization \cite{huber_1964,charbonnier_1997}. 
In comparison, the clinical software currently employed is a Feldkamp-Davis-Kress (FDK) reconstruction~\cite{fdk_1984}, an approximate filtered backprojection algorithm which, in its original form, only allows reconstruction of circular scan trajectories. While an FDK algorithm will remain limited in terms of scan trajectory compatibility, an iterative algorithm allows reconstruction of an arbitrary trajectory, enabling  direct comparison of CBCT volumes from different scan trajectories. 

\subsubsection{Evaluation Metrics}

To characterize the artifacts caused by the cone-beam geometry for each trajectory, the reconstructions of the phantom with disks are normalized with a reconstruction of a homogeneous phantom of the corresponding scan trajectory.
The magnitude of the cone-beam artifacts is assessed through the width $W$ of the axial disks. The average profile of the disk is computed over its axial surface, then its derivative is taken along the z-axis. The limits of the width of the artifact are defined by the z positions on each side of the disk where the absolute derivative of the profile reaches 1\% of the maximal absolute derivative.
This limit definition provides a width measurement of the complete profile based on its shape, rather than a threshold on the profile magnitude which may not reflect accurately the extent of the cone-beam artifact.
As the disk width $W$ is dependent on the geometry of the object, a scaled width $w$ is defined based on the thickness $t$ and the diameter $D$ of the disk, all in mm, with the general expression
\begin{equation}
    w = \frac{W-t}{D^{\alpha}} \text{\:\:\:,}
    \label{eq width normalization}
\end{equation}
where $\alpha$ is a constant established based on measurement of widths $W$ from simulations of various disk geometries. With this heuristic scaling, allowing disk-geometry-independent artifact evaluation, widths are measured for disks with diameters of 40, 30 and 20 mm and with thicknesses of 1 and 2 mm, positioned on the z-axis as well as shifted $\pm60$ mm along each axis in the axial plane. The width results are compared with the incompleteness values at the same position in the phantom to evaluate the accuracy of the incompleteness metric to estimate the magnitude of cone-beam artifacts.

%% file: Document/Resultats.tex
\section{Results} 

\subsection{Tomographic Incompleteness}

\begin{figure}[b!]
    \centering
    \includegraphics[width=\linewidth]{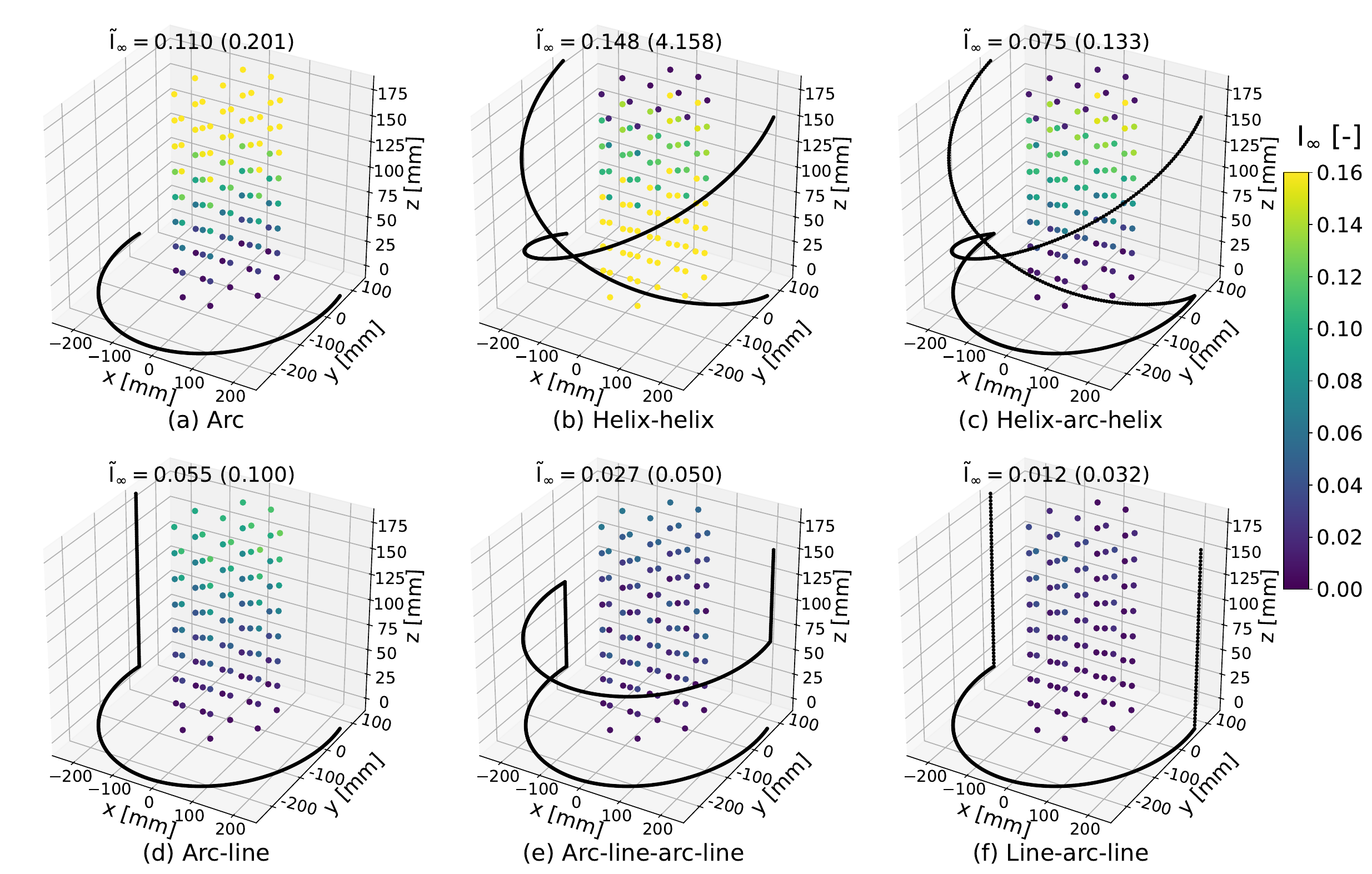}
    \caption{Tomographic incompleteness maps for different scan trajectories of the Gamma Knife's CBCT system, where lower is better. Median values $\tilde{\text{I}}_{\infty}$ of the distributions are written above each trajectory, with the 90th percentile within parenthesis. The black dots represent all discrete source positions during the scan, with the source-to-axis distance reduced by a factor of three. For improved visibility, only a subset of computed points are shown.}
    \label{fig: maps overall}
\end{figure}

Figure \ref{fig: maps overall} presents tomographic incompleteness maps for scan trajectories compatible with the Gamma Knife CBCT geometry, in which
(a) represents the incompleteness of the current single arc scan trajectory, where the angle $\psi$ increases linearly with the axial position. All the other trajectories, called exotic, are given an axial coverage of 175 mm, which is sufficient for a system dedicated to head scans. The median incompleteness across the FOV and its 90th percentile are written above each trajectory's incompleteness map.
A trajectory made of two opposite helices is shown in (b), and an arc is added to the double helices at the inferior edge of the scanned volume in (c) to create the trajectory helix-arc-helix. The comparison between these two trajectories indicates that, because of the piercing point on the inferior side of the detector, the current arc must be included in the new trajectory for sufficient sampling near the volume inferior edge. Also, a helical trajectory for a system with limited axial FOV and with a gantry limited to a 200$\degree$ range is deemed inadequate because of inefficient sampling.
Figure \ref{fig: maps overall}(d) shows the arc-line trajectory where a line with axial steps of the source of 1.75 mm is added after the current arc. 
A second arc is added in (e) around the center of the object to create the arc-line-arc-line trajectory, where each line segment retains the gantry angle from the last projection of the previous arc. Adding more arcs at other axial positions improves substantially sampling at higher axial positions, but requires many more projections, increasing the complexity of the trajectory. 
Finally, Figure \ref{fig: maps overall}(f) shows the line-arc-line trajectory where the lines before and after the arc are made with axial steps of the source of 3.5 mm. Adding lines at the first and last gantry angles provides much better volume sampling because of the near-opposite angles and the long axial coverage, while requiring a smaller increase in the number of projections. With a median incompleteness across the FOV of 0.012, a line-arc-line scan trajectory should, therefore, have very few visible cone-beam artifacts, as most incompleteness values inside the FOV are under 0.02.

\begin{figure}[htb!]
    \centering
    \includegraphics[width=\linewidth]{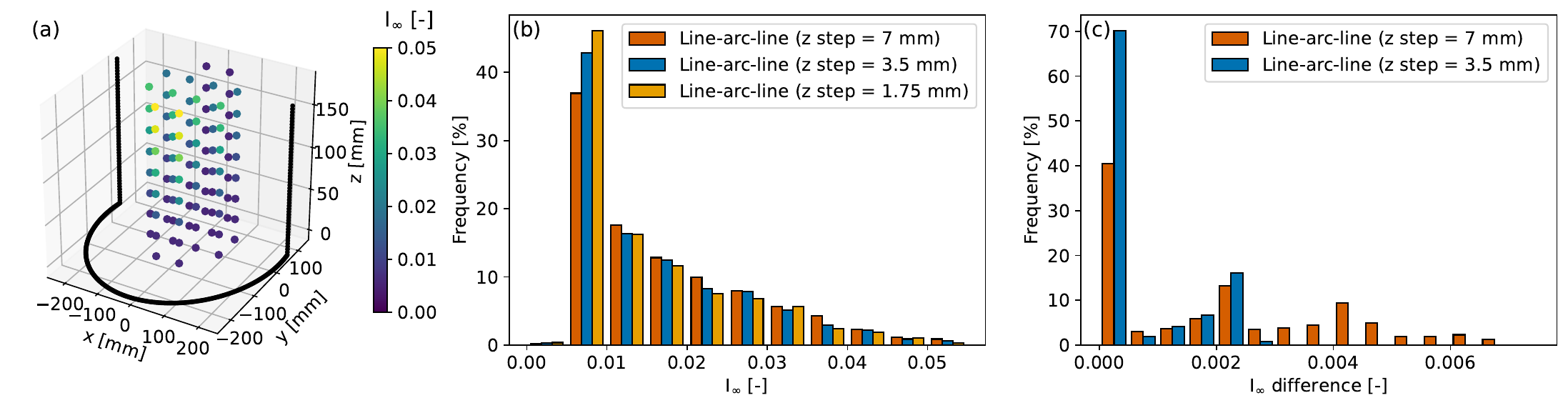}
    \caption{Tomographic incompleteness maps for the line-arc-line trajectories. (a) shows selected points of the $I_{\infty}$ map where 50 projections are acquired per line, with a 3.5-mm axial step. The black points represent the source trajectory, with the source-to-axis distance reduced by a factor of three for improved visibility. The histogram in (b) compares the distribution of $I_{\infty}$ values in the FOV for line-arc-line trajectories with different axial steps of the source during the line segments. The histogram in (c) shows the difference in the $I_{\infty}$ values in the FOV between the trajectories with axial steps of 3.5 and 7~mm and the one with a step of 1.75~mm.}
    \label{fig: maps lines}
\end{figure}

Figure \ref{fig: maps lines}(a) shows the incompleteness map of the line-arc-line trajectory with z-steps of 3.5 mm during the line segments, for a tighter visualization scale compared to Figure \ref{fig: maps overall}(f). The $I_{\infty}$ values above 0.02 are gathered in the anterio-superior region. 
The histograms in the panels (b,c) of Figure \ref{fig: maps lines} show the effect of the axial step between each source position along a line trajectory. In (b), while the incompleteness distribution is similar for each source step, when the axial step is larger, there is a decrease in the proportion of incompleteness values below 0.01, resulting in an increase of higher values. The proportion of incompleteness values above 0.02 increases from 26\% to 28\% and 32\% when the source step is increased from 1.75 mm to 3.5 and 7 mm, respectively. 
The histogram in (c) highlights the individual increase of incompleteness (incompleteness difference) at each point computed in the FOV when the axial step is larger than 1.75 mm. For a step of 3.5 mm, most incompleteness values increase by less than 0.0005, and the increase remains below 0.003. However, for a source step of 7 mm, the variation in incompleteness rises up to 0.007, representing an increase of 50\% compared to the median incompleteness of the line-arc-line trajectory with z-steps of 1.75 mm. 
Consequently, the line-arc-line scan trajectory with an axial step of 3.5 mm seems the most promising for improved image quality while requiring an increase of the number of projections of only 30\%.

Since a discrete number of plane normals are sampled, the reported incompleteness values are lower bounds of the true incompleteness. An upper bound can be estimated from the cone angle between sampled plane normals. With the use of 200,001 planes uniformly distributed on a hemisphere to establish the most undersampled direction, the half-cone angle between points is approximately 0.17$\degree$, leading to an upper bound for the estimation of the incompleteness value of $+0.003$ for $I_{\infty}$ below 0.4, and a higher upper bound for $I_{\infty}$ above 0.4.

\subsection{CBCT Simulations}

The volumes generated by the FDK and WLS reconstruction algorithms for the same single-arc scan trajectory are shown in Figure \ref{fig: disk all trajectories}(a,b) of the Appendix. The algorithms produce comparable CBCT volumes, with similar magnitude of cone-beam artifacts, validating the use of the WLS reconstruction for this study.

\begin{figure}[htb!]
    \centering
    \includegraphics[width=0.87\linewidth]{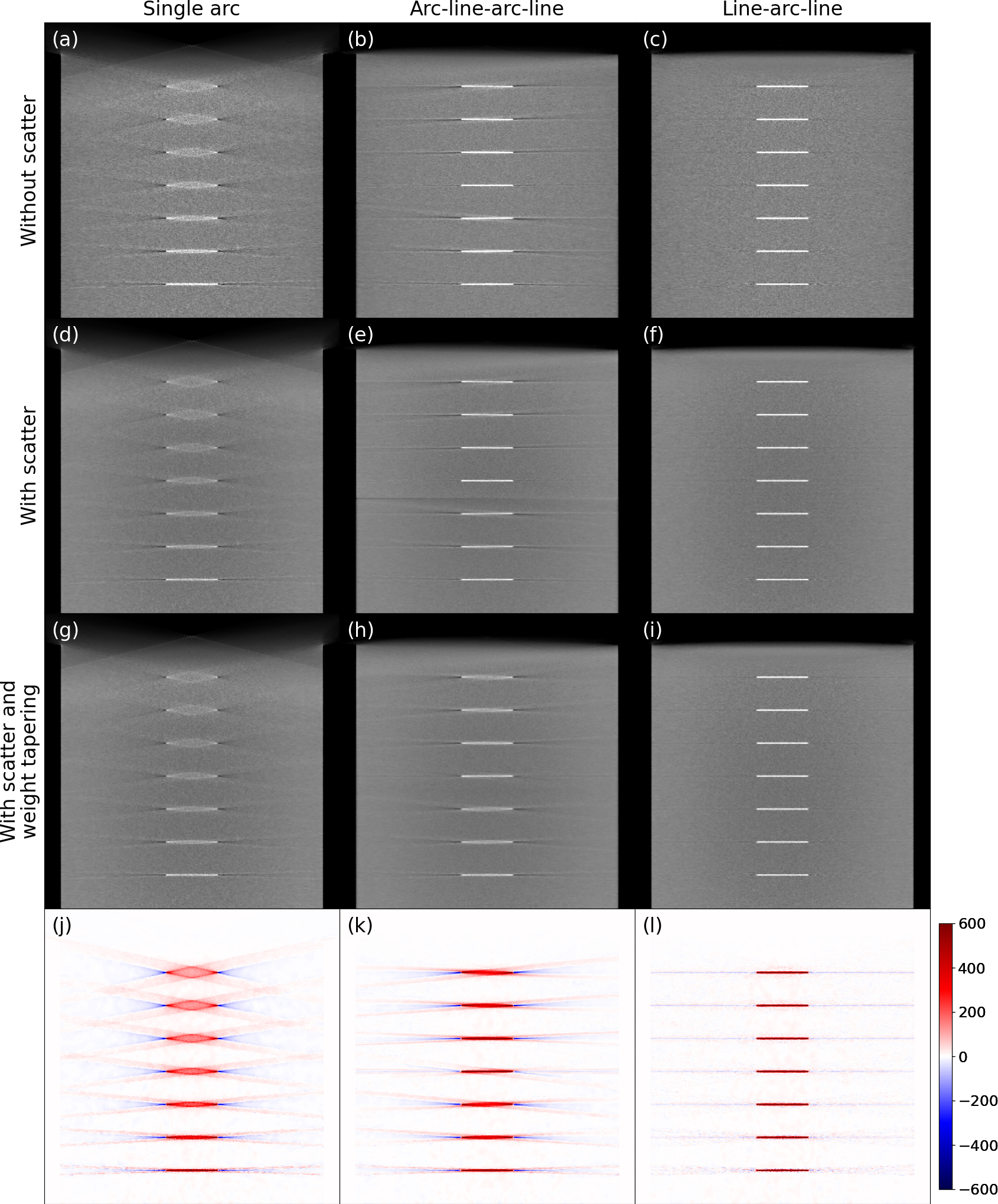}
    \caption{Coronal slices from the CBCT volumes of the virtual phantom with unit disks using the iterative reconstruction, for the trajectories (a,d,g) single arc, (b,e,h) arc-line-arc-line with source axial steps of 1.75 mm and the second arc at 87.5 mm in axial, and (c,f,i) line-arc-line with source axial steps of 3.5 mm. The first row shows simulations without scatter, the second row simulations with scatter and the third row simulations with scatter and weight tapering to the inferior edge of projections (window: [-1000, 1000] HU). The last row images (j,k,l) correspond to the difference between each image in (g,h,i) and the corresponding CBCT image of a homogeneous phantom.}
    \vspace{5mm}
    \label{fig: disk overall scatter}
\end{figure}

Figure \ref{fig: disk overall scatter} presents CBCT slices of the virtual phantom with centered unit disks, for selected scan trajectories and reconstructed with the WLS iterative algorithm. The first and second rows are volumes from simulations without and with scatter, respectively. For the single-arc scan, shown in the first column, the cone-beam artifacts created by the axial disks are increasing in the superior direction, as seen by the distortion of the disks. Among the new trajectories investigated, the line-arc-line one produces the reconstructions with the smallest cone-beam artifacts: these artifacts are, in fact, virtually gone for all disks positioned on the central axis.

As suggested by the incompleteness maps, adding a second arc at a higher axial position is not the most efficient to reduce the overall cone-beam artifacts. CBCT simulations with scatter, shown in the second row of Figure \ref{fig: disk overall scatter}, reveal that, in addition, there is a new artifact right under the source's axial position for the second arc. This artifact is caused by the difference in scatter magnitude when the X-ray beam propagates through a different section of the phantom, leading to a lower scatter magnitude in the projections acquired at a more superior position, compared to the projections acquired at the inferior base of the phantom. Consequently, to compensate for the discrepancy in intensity between overlapping projections caused by different amounts of scatter, the iterative reconstruction algorithm tends to assign lower attenuation values to voxels inferior to the overlapping region. This effect is particularly prominent for the trajectory with an additional arc for two main reasons. First, the source is moved almost half the length of the phantom, reducing the beam path through the phantom, which significantly reduces the amount of scatter. Second, many projections are acquired at the same axial position, leading to groups of projections with similar scatter magnitude for each arc. The iterative algorithm is then pushed to consider the intensities from both arcs with a similar weight, creating the severe artifact in the reconstruction as a compromise. Even CBCT simulations with a sparsely sampled second arc have the artifact with a similar intensity as the simulations with a regularly sampled second arc.
CBCT simulations show that the artifact, called here the scatter discrepancy artifact, is also in the CBCT volumes from the line-arc-line trajectory for each source axial step, but they are more subtle. 

To remove this artifact, during the iterative reconstruction, a linear weight tapering is applied to the intensities from the inferior edge of the projection, up to a range proportional to the axial step. For simulations with a second arc, the tapering range is proportional to the distance between the arcs. Simulations have established that a range defined by 120\% of the step is necessary to remove most of the artifacts.
The results are presented in the third row of Figure~\ref{fig: disk overall scatter}. This correction can efficiently remove the scatter discrepancy artifact. However, the lower weighting of some relevant data leads to a small degradation of image quality in terms of other artifacts, such as cone-beam artifacts and exponential edge gradient effect (EEGE). The latter is a dark streak along the direction of a straight and sharp high-density edge. 
It is caused by data interpolation failure during reconstruction when the non-linear beam attenuation, through a high-density straight edge in the axial plane, is spatially averaged over a detector cell \cite{eege_1981,Schulze_eege_2011,Schwarz_eege_2011}.
As the line-arc-line simulations require tapering over a short range only, the resulting volume still has no visible cone-beam artifacts, although the EEGE is more visible. On the other hand, the wide tapering required for the arc-line-arc-line simulations causes a visual increase of the cone-beam artifacts. This further demonstrates the advantages of the line-arc-line trajectory compared to a trajectory with a second arc.

The last row of Figure \ref{fig: disk overall scatter} shows the artifacts created by the disks for each trajectory, by means of the difference between the reconstruction of the virtual phantom with disks and an identical reconstruction of a homogeneous phantom, with scatter and weight tapering included in the simulations. 
Figures \ref{fig: disk overall scatter}(i) and (l) from the line-arc-line trajectory show the most improvement in terms of cone-beam artifact reduction and improved out-of-FOV reconstruction accuracy at the superior extremity, beyond the most superior disk. Reconstructions from more trajectories investigated are shown in Figure \ref{fig: disk all trajectories} of the Appendix. The axial step of the source during the line movement is an optimization parameter that will have an effect on imaging time, patient dose and image quality. While steps of 7 mm seem to be sufficient to remove all major cone-beam artifacts, smaller steps allow for sharper images with fewer artifacts caused by the EEGE. The EEGE is actually similar to the partial volume effect in helical CT, resulting from undersampling in the z-direction during multi-slice acquisition \cite{Schulze_eege_2011,Marshall_eege_2022}. With shorter z-steps of the source, the reconstruction algorithm can better interpolate data, particularly sharp high-density edges which align with the beam direction for a certain z-position of the source, where a non-linear beam attenuation effect occurs on the detector.

\begin{figure}[htb!]
    \centering
    \includegraphics[width=\linewidth, trim={0 0 3.9cm 0},clip]{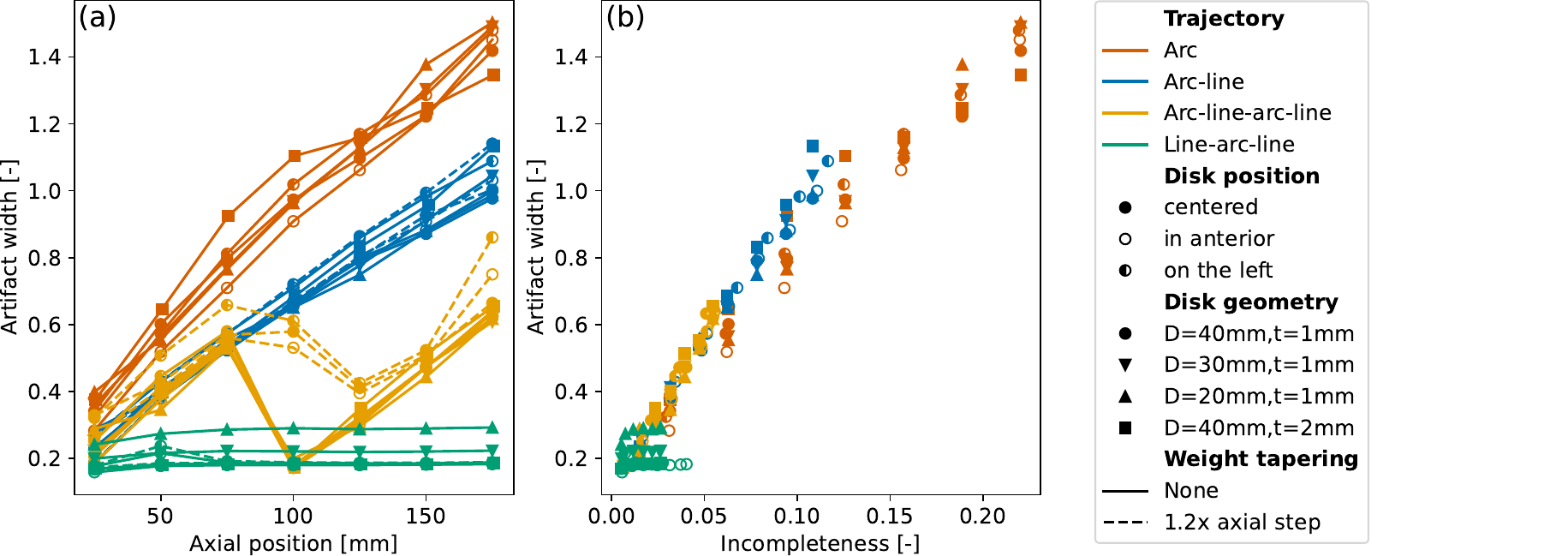}
    \caption{Geometry-independent widths of the cone-beam artifacts created by disks, measured at 1\% of the maximal derivative of the profile created by the disks, plotted against (a) the axial position of the disk and (b) the incompleteness value associated with that disk position and scan trajectory.}
    \label{fig: width metric}
\end{figure}

From the normalized reconstructions, shown in the last row of Figure \ref{fig: disk overall scatter}, the width of the cone-beam artifact is measured. From simulations with disks of different $t$ and $D$, the constant $\alpha$ from equation \ref{eq width normalization} for the disk geometry normalization is calculated as 0.7. Figure \ref{fig: width metric}(a) shows the scaled widths of disks according to their axial position in the phantom, for selected trajectories and different disk geometries and positions in the axial plane. While the use of weight tapering during reconstruction barely affects the widths from trajectories with only one arc, those from the trajectory with two arcs are significantly increased, particularly the widths slightly above the second arc. Excluding reconstructions with weight tapering, the artifact width for a given axial position varies mainly depending on the scan trajectory. The single arc trajectory has the highest widths, which increases linearly with the axial position. The widths from the arc-line simulation also increase linearly with the axial position, but they remain lower than those from the single arc simulation. The arc-line and the arc-line-arc-line have very similar results for the first three disks from the inferior side, as the sampling is the same at these axial positions. The second arc of the more complex trajectory then reduces the cone-beam artifacts for the subsequent disks. Finally, the line-arc-line trajectory has constant low widths for every disk, as no cone-beam artifacts are visible in the reconstructions, regardless of the disk position.

Figure \ref{fig: width metric}(b) compares the widths measured for every disk from reconstructions without weight tapering, with the $I_{\infty}$ values associated with each point in the volume and the scan trajectory. There appears to be a strong correlation between the two measurements, independent of the scan trajectory and the disk geometry and position. A higher incompleteness should lead to a stronger cone-beam artifact, if the point of interest in the phantom is on a structure susceptible to such artifacts. Consequently, the incompleteness values, obtained by geometrical analysis, seem to be representative of the magnitude of the cone-beam artifacts produced by each scan trajectory investigated, before considering the application of weight tapering during reconstruction.

\begin{figure}[htb!]
    \centering
    \includegraphics[width=1\linewidth]{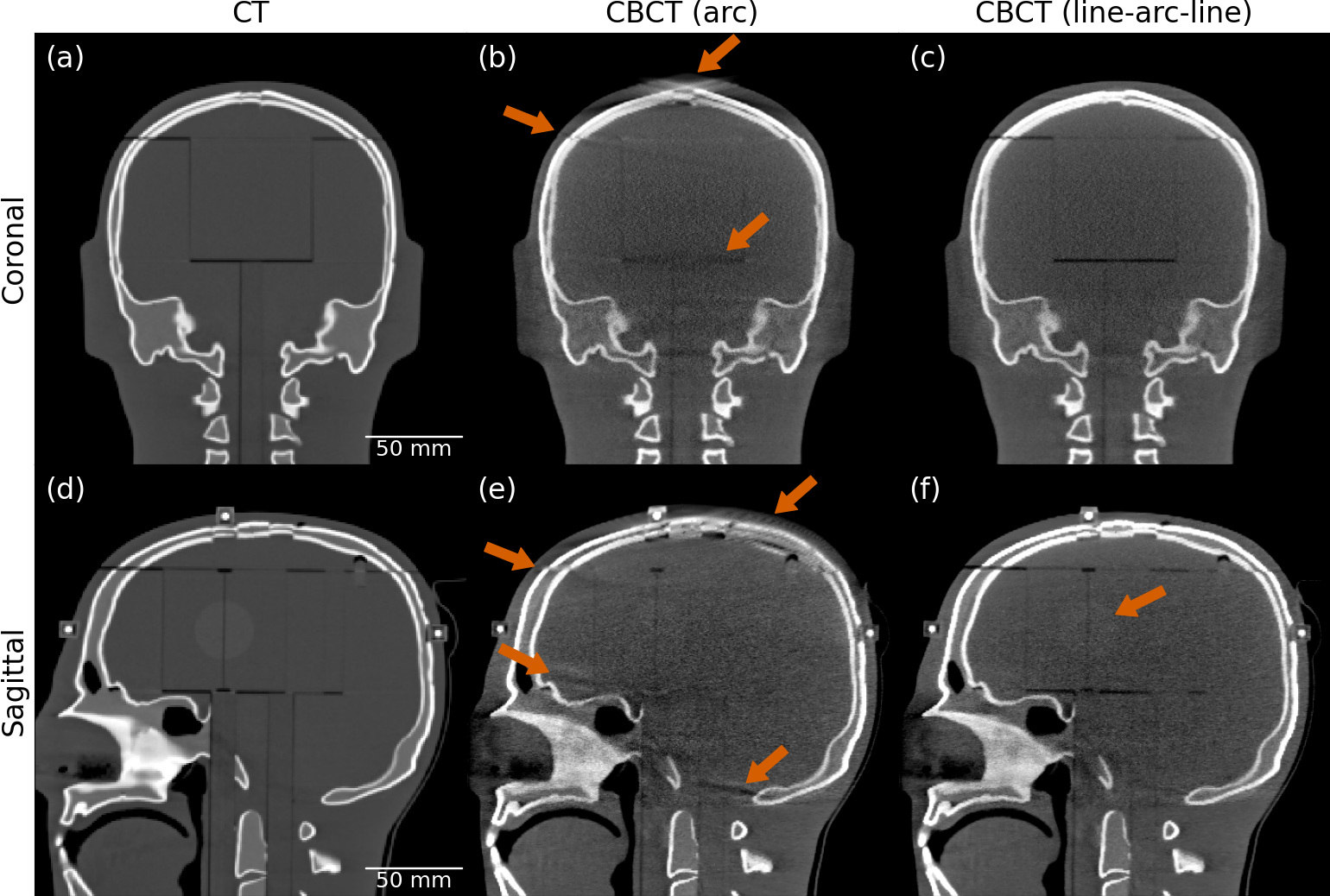}
    \caption{Slices of the CT image of the head phantom in the first column, and the corresponding slices of the CBCT images, for the current single arc scan trajectory in the second column compared to the proposed line-arc-line trajectory with source axial steps of 3.5 mm in the last column, reconstructed with the iterative algorithm (window: [-500, 1500] HU). Arrows indicate relevant structures affected by the change of scan trajectory.}
    \label{fig: steev best vs arc wls vs ct}
\end{figure}

Finally, Figure \ref{fig: steev best vs arc wls vs ct} shows the reconstruction from the head phantom for a line-arc-line scan, compared to the single arc scan. Since no scatter discrepancy artifact can be seen in the reconstruction, no weight tapering is applied to the projections during reconstruction. 
The difference map in the last column indicates most improvements are at the superior edge of the skull, but there is also an overall better uniformity inside the brain, and the axial junctions of the different pieces of the phantom are much better defined. The low contrast sphere inside the phantom visible in the sagittal view is also visualized better with the line-arc-line trajectory.

%% file: Document/Discussion.tex
\section{Discussion} 

To establish the optimal scan trajectory for the Gamma Knife CBCT, two approaches, which are tomographic incompleteness maps and CBCT simulations, have been followed and reach the same conclusion: out of the investigated trajectories, a line-arc-line trajectory is the best for improving the image quality while limiting the complexity and the amount of additional projections. 

\subsection{Tomographic Incompleteness}

The tomographic incompleteness maps provide a theoretical approach to establish at any point in the volume the level of missing data for exact reconstruction in the most undersampled direction, based on the CBCT geometry and the scan trajectory. This calculation demonstrated how the arc at the inferior edge of the phantom is necessary for the Gamma Knife CBCT system with its piercing point at the inferior edge of the detector. The incompleteness maps also show that a complex trajectory made of helices is inefficient in the context of a gantry with a limited angular range; lines and arcs provide better sampling and are easier to implement in practice. Adding more arcs only improves sampling above each arc and the effect decreases linearly with the distance, leading to a non-uniform incompleteness map with very low values just above the arcs but higher values under the arcs, which may lead to cone-beam artifacts. As each arc adds a significant number of projections to the trajectory, having a trajectory with more than one arc is not optimal. Even if CBCT simulations with a sparsely sampled second arc revealed similar image quality to a simulation with a regular sampled second arc, as shown in Figure \ref{fig: disk all trajectories}(e,f) of the Appendix, such a trajectory provides limited sampling improvement. 

However, adding line trajectories improves the sampling across the complete axial range of interest with few additional projections. With the gantry angle range of 200$\degree$, lines at the first and last angles are easy to integrate into the scan workflow and they provide near-opposite sampling. 
This study also demonstrates the importance of the choice of the source position for the new projections. For instance, when 100 additional projections are distributed along two lines instead of only one, the incompleteness distribution shows a reduction of 78\% in the median incompleteness, with a reduction of 68\% in the 90th percentile. 

Since the line-arc-line trajectory requires less gantry movement compared to the helix-arc-helix and arc-line-arc-line trajectories, it is expected to limit the additional scan time and allow for easier integration into the workflow.
It may also limit imaging dose, since the number of additional projections is small for a significant sampling improvement. It would be interesting to assess whether the dose per projection could be lowered so that the total patient dose is the same as, or even lower than, the current arc scan, while maintaining an improved image quality. With the cone-beam artifacts caused by geometry, a lower exposure should not affect the improvements seen in this work. However, for a given trajectory, reducing the imaging dose by a given factor should reduce the signal-to-noise ratio by approximately the square root of that dose factor \cite{mao_improvement_2018}. Overall, when comparing different trajectories giving the same imaging dose, the quantum noise should be similar, but a small increase in electronic noise is expected for trajectories with a higher number of projections \cite{zhao_noise_2014}.

The histograms of Figure \ref{fig: maps lines} indicate that the choice of the axial step of the source in the line segment does not considerably affect the range of incompleteness values in the FOV. It mostly affects the repartition of values, with fewer values below 0.01 when the source step is increased.
When compared to the plane computation upper bound of +0.003, a step of up to 3.5 mm has a negligible difference in incompleteness values with a maximal increase of 0.003 compared to a step of 1.75 mm. Therefore, these 3.5-mm steps across a line of 175 mm provides a good trade-off between image quality, on one hand, and patient dose and scan time, on the other.

Overall, the line-arc-line trajectory has a median incompleteness value of 0.012 with lines made of only 50 projections. Most points in the FOV have incompleteness values below 0.02: as shown in Figure \ref{fig: maps lines}(a), only the anterosuperior region has higher incompleteness values due to the arc at the inferior edge reaching the limit of its FOV. As cone-beam artifacts are object dependent, the presence of some small cone-beam artifacts in the anterosuperior region with the line-arc-line depends on the object of interest. 

According to the tomographic incompleteness analysis, the sampling with the line-arc-line scan trajectory remains mostly favourable in the volume of interest, which is a head. Beyond a radius of 100 mm and a height of 180 mm, more visible artifacts may appear. The limits of the improvement in the axial direction are not defined by the length of the line, but rather by the FOV of the arc at the inferior edge. Extending the FOV beyond 180 mm in superior would require sampling at more angles at a higher axial position, such as with a second arc.

While the incompleteness values are obtained through the analysis of 200,001 uniformly distributed planes on a hemisphere, there is still some underestimation. The upper bound to the incompleteness values indicates the limit of the estimation accuracy. When comparing scan trajectories with significant sampling differences, such as the trajectories of Figure \ref{fig: maps overall}, a smaller number of planes is sufficient to differentiate the level of sampling of each trajectory. However, the limit of the estimation can be reached when comparing similar trajectories, such as the line-arc-line ones of Figure \ref{fig: maps lines} which vary only by the choice of the source axial step. 
Also, the better the sampling of the trajectories, the more planes are required to differentiate them. 
Consequently, the analysis of the line-arc-line trajectories reaches its limit of accuracy when comparing the results with the axial steps of 1.75 and 3.5 mm.

\subsection{CBCT Simulations}

Cone-beam artifacts are oriented in the perpendicular direction of the most undersampled plane, which corresponds to the axial plane for an arc scan trajectory. A large density variation along that plane will create cone-beam artifacts if the volume is undersampled. CBCT simulations of a virtual phantom with bone-equivalent disks aligned along the axial direction present cone-beam artifacts with a magnitude dependent on the position, the scan trajectory and the disk geometry. 

As shown in Figure \ref{fig: disk overall scatter}, simulations without scatter present no visible cone-beam artifacts and some slight EEGE artifacts in the reconstructions of the line-arc-line trajectory, whereas the arc-line-arc-line trajectory has some small cone-beam artifacts increasing with the axial position above each arc and some more visible streaks. The latter is caused by the single-side lines, similar to the streaks from the arc-line trajectory, in the Figure \ref{fig: disk all trajectories}(c) of the Appendix. While the incompleteness doesn't estimate EEGE and its streaks specifically, the simulations show a clear benefit in having two lines at near opposite gantry angles instead of just one, both in terms of cone-beam artifacts and streaks. 

The addition of scatter in the simulations also revealed a new artifact in an axial plane other than the most inferior one, whose severity increases with the proportion of projections acquired with the source at that axial position. This scatter discrepancy artifact could not have been predicted by the geometrical incompleteness metric. The artifact is subtle for every step along a line trajectory, but creates a clear streak and step of intensity when an arc is added above the inferior edge. Adding lines amid the trajectory with multiple arcs seems to reduce the severity of the artifact, and allows easier removal of the artifact with weight tapering of the shifted projections, as shown in the comparison of the arc-arc and arc-line-arc-line trajectories in Figure~\ref{fig: disk all trajectories}(d-e) respectively. The scatter discrepancy artifacts from both the simulations of the arc-line-arc-line and line-arc-line trajectories are mostly removed with the weight tapering. However, the longer range along which tapering is necessary for the trajectory with a second arc causes loss of relevant information for artifact-free reconstruction. 
For 40-mm diameter and 1-mm thick disks, the artifact widths $W$ from the line-arc-line trajectory only increase on average by 1.8\% when tapering is applied, whereas the ones from the line-arc-line-arc increase on average by 37\%, with a maximal increase of 176\%.

The appearance of a new artifact due to scatter when using an exotic scan trajectory reinforces the relevance of developing an accurate scatter estimate to remove scatter from the projections. 
Standard scatter effects include cupping, streaks, lower soft-tissue contrast and underestimation of structure density \cite{Schulze_eege_2011,Schwarz_eege_2011}. The scatter discrepancy artifact is now added to that list which would benefit from scatter removal. In the context of choosing between exotic scan trajectories, one strategy could be selecting one that is less reliant on scatter correction by means of less prominent scatter discrepancy artifacts.
Nevertheless, in this study, without an accurate scatter estimate which could remove all scatter artifacts, results from both the simulations with and without scatter indicate that adding two near-opposed lines and no additional arc to the current single-arc trajectory is the most promising among the investigated trajectories. 

The effect of the axial step size of the source during the line segment can mostly be visualized in terms of EEGE artifacts, as in Figure \ref{fig: disk all trajectories}(g-i) of the Appendix. The shorter the step, the less prominent the streaks from EEGE. The choice of the step is then a matter of compromise between image quality, scan time and patient dose. 
Since the EEGE is not a direct cause of geometric sampling incompleteness, it should be noted that incompleteness maps cannot be used to predict these artifacts.
Furthermore, the streaks from EEGE created by the disks in the virtual phantom constitute an extreme case of the artifact, with the perfectly sharp and long straight edges of the disks. A typical object scanned such as a head does not have such edges, therefore less EEGE affects the reconstruction. For instance, no streaks attributed to EEGE are present in the head phantom reconstruction of the line-arc-line trajectory of Figure~\ref{fig: steev best vs arc wls vs ct}(c,f). 

To quantify the cone-beam artifacts in the reconstructions, the widths of the disks were evaluated with a heuristic metric based on the results from trajectories composed of arcs and lines. 
Measuring the artifact width at 1\% of the maximal derivative ensures consideration of the whole artifact, based on the shape of the profile. For instance, some trajectories such as arc-line produce artifacts with intensities only below half of the profile's maximum intensity. Well-sampled trajectories like the line-arc-line can have a peak profile with a small remaining increased background near the edge of the disks. Therefore, measuring a width defined by an intensity threshold is harder to generalize to any scan trajectory, but using a very low derivative threshold seems to capture well the extent of the cone-beam artifact.


Comparing the normalized widths with the corresponding incompleteness values show a strong correlation independent of the scan trajectory. A theoretical analysis of the potential of a scan trajectory through incompleteness maps is therefore adequate to estimate the magnitude of cone-beam artifacts. However, as scatter can create a new discrepancy artifact for exotic trajectories and some correction may be necessary such as weight tapering on the projections during reconstruction, the magnitude of the cone-beam artifacts may be greater than the estimate from the incompleteness value. An exotic trajectory made of multiple arcs will be much more affected than one made only of lines added to the standard arc. 
Consequently, while the incompleteness metric indicates the amount of cone-beam artifacts that can be expected due to the geometry, the incompleteness map will underestimate the cone-beam artifacts in the final volume if the scan trajectory has multiple arcs.
Both of these metrics also fail to evaluate EEGE, which leaves a need for visual analysis of the reconstructions with trajectories including line segments.

The CBCT reconstructions of a head phantom of Figure \ref{fig: steev best vs arc wls vs ct} are consistent with their respective scan trajectory. Compared to the single arc, the line-arc-line trajectory produces reconstructions with noticeable less cone-beam artifacts, mostly visible on the superior edge of the skull, but also at the axial junctions of the various assembled pieces of the phantom. There is also better uniformity inside the brain, mostly in homogeneous areas near sharp density changes, which are visible in the sagittal plane where the sampling is limited by the gantry angular range. This uniformity improvement also helps with soft-tissue contrast, for instance with the low contrast sphere visible in the brain region of the head phantom. The absence of weight tapering for the reconstruction of the line-arc-line simulation indicates that the scatter discrepancy artifact has become non-noticeable with a more complex phantom such as a head, compared to the mostly homogeneous phantom with disks. The increase of the EEGE, associated with the application of weight tapering, is therefore avoided with a head phantom. 

In short, this study demonstrates the theoretical potential of the line-arc-line trajectory as a new scan trajectory for the Gamma Knife CBCT. Experiments conducted on a research Gamma Knife are needed to confirm the feasibility of the proposed line-arc-line trajectory and to assess the real image quality improvement.

%% file: Document/Conclu.tex
\section{Conclusion}

As the single-arc scan trajectory does not provide sufficient sampling for geometric artifact-free reconstructions, a new trajectory was proposed for the Gamma Knife CBCT, based on tomographic incompleteness analysis as well as CBCT simulations of a virtual phantom with disks aligned along the z-axis and of a head phantom. For both approaches, a line-arc-line scan trajectory seems the most promising for reducing cone-beam artifacts while limiting the amount of additional projections to the current scan. Compared to a trajectory with multiple arcs, the proposed trajectory limits the magnitude of a secondary artifact caused by a variation in scatter magnitude when the source is axially translated. Choosing a shorter axial step of the source during the line segments may increase scan time and patient dose, however the streaks caused by EEGE will be reduced. 
Overall, the line-arc-line scan trajectory has the potential to remove most cone-beam artifacts in Gamma Knife CBCT images while adding a patient table translation before and after the arc scan seems reasonable to integrate into the current imaging workflow.

%% file: Document/Annexe.tex
\label{appendix}

\begin{table}[htb!]
    \centering
    \caption{RECORDS checklist (TG-268) for Monte Carlo simulations.}
    \begin{tabular}{lp{2.9cm}p{7.5cm}p{2.4cm}}
        \hline
        \# & Item & Description & References  \\
        \hline
        2,3 & Code, version/release & Elekta internal software, based on GPUMCI (GPU Monte Carlo simulation for Imaging) & \cite{gpumci_2017} \\
        4, 17 & Validation & GPUMCI previously validated against EGSnrc and GPUMC using Elekta Synergy CBCT system parameters & \cite{gpumci_2017}  \\
        5 & Computation time &  On a NVIDIA RTX A2000 GPU, from a CT reconstruction, it takes about 2-3h to simulate 350 to 550 projections, each made from about $10^9$ histories  &  \\
        8 & Source description & Monte Carlo-generated phase space provided by Elekta, from an X-ray source of 90 kVp positioned at 79.0 cm from the central axis, after collimation and bowtie filter  &  \\
        9 & Cross sections & Compton, Rayleigh and photoelectric interactions from xraylib library: Photoelectric attenuation cross sections (assuming photons absorbed in place) through mass attenuation coefficients; Rayleigh and Compton interaction cross sections in pre-computed lookup tables.
         &  \cite{gpumci_2017,xraylib} \\
        10 & Transport parameters & Electrons ignored as they don't contribute to the image formation with their negligeable penetration depth. Photons have a cut-off energy of 10 keV, their step length is decided by Woodcock-delta tracking. & \cite{woodcock} \\
        11 & VRT/AEIT & No variance reduction techniques used &  \\
        12 & Scored quantities & Primary and scatter reaching a energy integrating flat panel detector &  \\
        13, 18 & Histories/uncertainties & 60,424,502 photons re-used 16 times, giving a total of about $10^9$ simulated histories &  \\
        14 & Statistical methods & No statistical uncertainty for this work, but previous assessment of relative error in energy fluence images between GPUMCI and EGSnrc showed a relative error of 0.4\% in the primary and of 1.1\% in the scatter. & \cite{gpumci_2017} \\
        15, 16 & Post-processing & Scored quantities are spatially denoised with a bidimensional Gaussian filter with a variance of 4~mm &  \\
        \hline
    \end{tabular}
    \label{tab: tg268}
\end{table}

\begin{figure}[htb!]
    \centering
    \includegraphics[width=\linewidth]{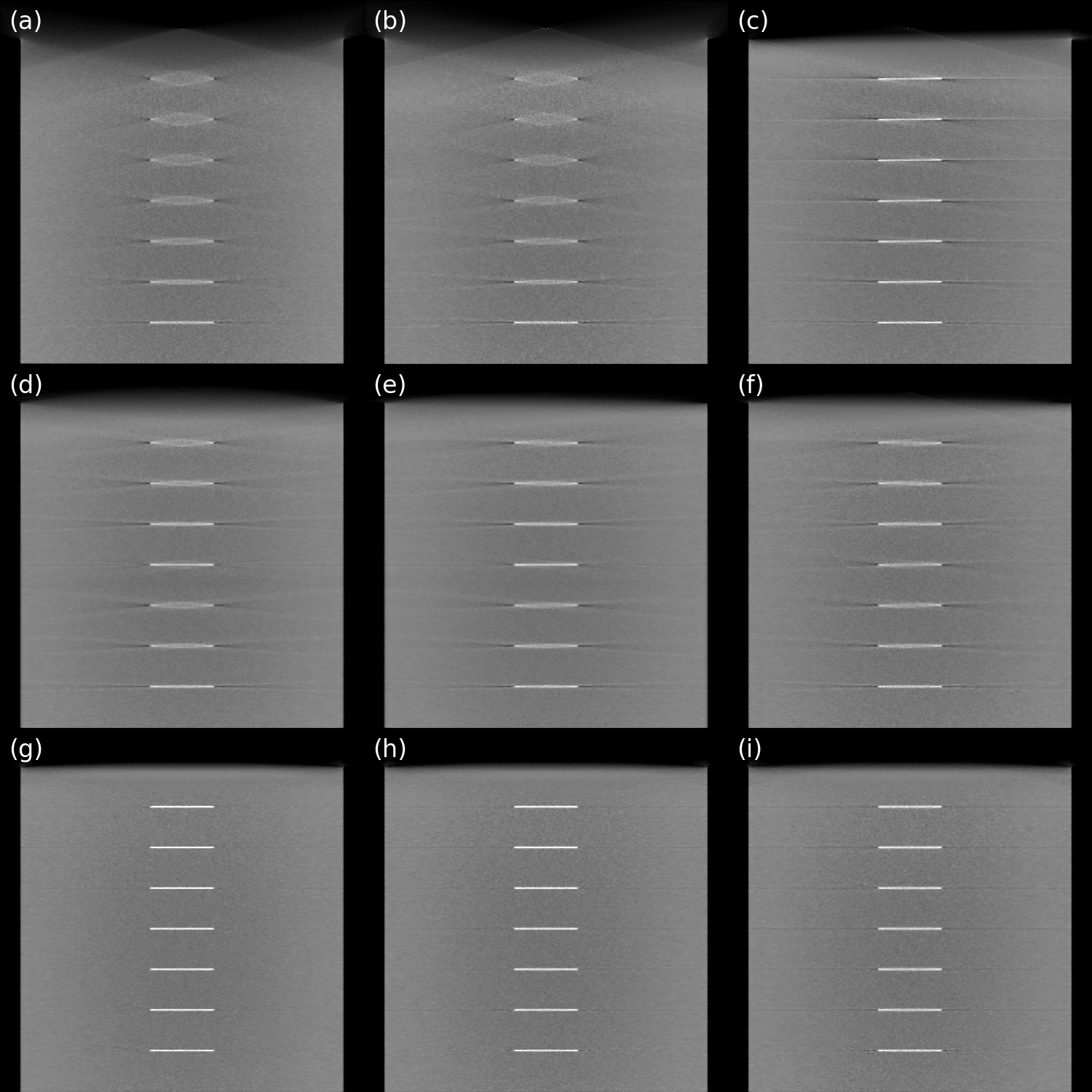}
    \caption{Coronal slices of the CBCT volumes of the virtual phantom with disks, produced with (a) FDK reconstruction and (b-i) WLS reconstruction. The scan trajectories are (a,b) single arc, (c) arc-line with source axial steps of 1.75 mm, (d) arc-arc with the second arc at 87.5 mm in axial, (e) arc-line-arc-line with source axial steps of 1.75 mm and the second arc at 87.5 mm in axial, (f) arc-line-arc-line which differs from (e) with the second arc made with 10\% of the projection number in the first arc, and (g,h,i) line-arc-line with source axial steps of 1.75 mm, 3.5 mm and 7 mm, respectively (window: [-1000, 1000] HU). Weight tapering is applied during iterative reconstruction.}
    \label{fig: disk all trajectories}
\end{figure}


\newpage